\documentstyle[aps,prl,epsf,floats]{revtex}
\def\figuresize{\ifpreprintsty 12cm \else 8cm \fi}

\begin{document}

\ifpreprintsty
\else \twocolumn[\hsize\textwidth\columnwidth\hsize\csname@twocolumnfalse%
\endcsname \fi

\draft
\title{Do interactions increase or reduce the conductance
       of disordered electrons? \\It depends!}

\author{Thomas Vojta, Frank Epperlein and Michael Schreiber}
\address{Institut f\"ur Physik, Technische Universit\"at, D-09107 Chemnitz, Germany}
\date{version August 18, compiled \today}
\maketitle

\begin{abstract}
We investigate the influence of electron-electron interactions on the conductance of
two-dimensional disordered spinless electrons.
We present an efficient numerical method based on diagonalization in a
truncated basis of Hartree-Fock states to determine with high accuracy
the low-energy properties in the entire parameter space. 
We find that  weak interactions {\em increase} the d.c.\ conductance in the strongly localized 
regime while they {\em decrease} the d.c.\ conductance for weak disorder. Strong 
interactions always decrease the conductance. 
We also study the localization of single-particle excitations at the Fermi energy 
which turns out to be only weakly influenced by the interactions.
\end{abstract}


\ifpreprintsty
\else ]  \fi           

The influence of electron-electron interactions on the transport in disordered 
electronic systems has been investigated intensively within the last two decades
\cite{85rev,belitzrev94}.  Recently, the problem has reattracted a lot of attention 
after experimental \cite{2DMIT} and theoretical \cite{TIP} results challenged established opinions.  

It is well accepted \cite{kramrev93} that non-interacting electrons in three
dimensions (3D) undergo a localization-delocalization transition at finite
disorder. In contrast,
all states are localized  in 2D and 1D even for infinitesimal weak disorder \cite{SO}.
However, today it is believed that the metal-insulator transition (MIT) in most 
experimental systems cannot be explained based on non-interacting 
electrons.
The metallic phase of disordered interacting electrons has been studied intensively 
within the perturbative renormalization group (RG) \cite{belitzrev94} leading
to a qualitative analysis of the MIT and the identification of different universality 
classes. One of the results is that the lower critical dimension of the MIT is $d_c^-=2$ as it is
for non-interacting electrons. Therefore it came as a surprise when experiments \cite{2DMIT} 
on Si-MOSFETs revealed indications of a MIT in 2D. Since these experiments
are performed at low electron density where the Coulomb interaction is particularly strong 
compared to the Fermi energy,
interaction effects are a likely reason for this MIT. A complete understanding
has, however, not yet been obtained. Explanations were suggested
based on the perturbative RG \cite{runaway}, non-perturbative effects 
\cite{nonperturb}, or  the transition being a superconductor-insulator transition rather than a MIT
\cite{SIT}. 

Theoretically, surprising results have been obtained for just two 
interacting particles in the {\em insulating} regime \cite{TIP}. It was found that two particles 
can form a pair whose localization length is much larger than that of a single particle. 
Later an even larger delocalization was suggested for clusters of three or more 
particles \cite{MIP}. In the case of  a repulsive electron-electron 
these delocalized states have rather high energy, thus their relevance 
for the low-energy properties of a degenerate system is not clear. It has been argued 
that the many-particle problem can be reduced to a few interacting quasiparticles 
above the Fermi surface \cite{TIPquasi}. This is, however, only possible, if the interactions 
do not change the nature of the ground state. All in all, not even the qualitative influence 
of interactions is understood in the insulating regime.

We have numerically studied disordered 2D spinless electrons. 
Our calculations are summarized in Fig.\ \ref{fig:sigzero} which is the main result of 
this Letter.
\begin{figure}
  \epsfxsize=\figuresize
  \centerline{\epsffile{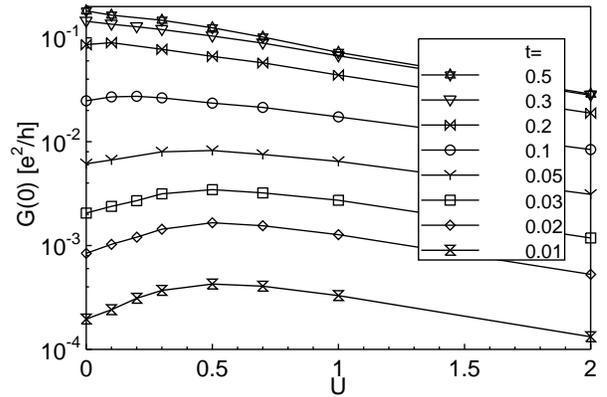}}
  \caption{d.c.\ conductance $G(0)$ for a system of $5^2$ sites as a function of interaction strength $U$
              for different kinetic energies $t$. The disorder is fixed at $W=1$. The statistical accuracy is 
              better than the symbol size.}
  \label{fig:sigzero}
\end{figure}
It  shows that the influence of repulsive electron-electron interactions on the d.c.\ 
conductance is opposite for high and low kinetic energies (i.e. weak vs. strong disorder). 
The conductance of strongly localized samples ($t=0.01$ to $0.03$) is considerably enhanced 
by a weak interaction. With increasing kinetic energy the relative enhancement
decreases as does the interaction range where the enhancement occurs. The conductance
of samples with the highest kinetic energies ($t=0.3$ and $0.5$) is reduced even 
by weak interactions.
In contrast, sufficiently strong interactions always reduce the conductance. This is not surprising since 
for large enough interaction strength the system will form a Wigner glass. 

These findings shed some light on seemingly contradicting numerical
results on the transport of disordered spinless electrons in the literature. 
Studies \cite{berkovits} of a 2D
model  in the {\em diffusive} regime yielded that interactions decrease  
the conductance. The same conclusion was drawn from density-matrix RG
studies \cite{dmrg_schmitt} and exact diagonalizations \cite{bouzerar} in 1D. 
In contrast, for 2D models in the {\em localized} regime
\cite{efros95,talamantes96} it was found that interactions lead to a delocalization.
Up to now it has been unclear whether these inconsistent results 
are due to  being in different parameter regions (weak vs.\ strong disorder), 
different quantities studied (conductance, many-particle level statistics 
or charge stiffness), or long-range vs.\ short-range interactions.
The results of this Letter suggest that being in different parameter regions 
is the most likely reason for the differences between the results cited above.
A result similar to ours was obtained recently \cite{fermimott} in a study of the 
ground state phase sensitivity in 1D.
It was found that for small disorder repulsive (nearest-neighbor) interactions reduce the phase sensitivity 
while for large disorder 
the phase sensitivity shows pronounced peaks at certain values
of the interaction.

In the remainder of the Letter we explain the model and the calculational method and further
discuss the results.
We consider a 2D quantum Coulomb glass model \cite{efros95,talamantes96,epper_hf,epper_exact}.
It is defined on a square lattice with $M=L^2$ 
sites occupied by $N=K M$ spinless electrons ($0\!<\!K\!<\!1$). To ensure charge neutrality
each site carries a compensating charge of  $Ke$. The Hamiltonian
reads
\ifpreprintsty
\begin{equation}
H =  -t  \sum_{\langle ij\rangle} (c_i^\dagger c_j + c_j^\dagger c_i) +
       \sum_i \varphi_i  n_i + \frac{1}{2}\sum_{i\not=j}(n_i-K)(n_j-K)U_{ij},
\label{eq:Hamiltonian}
\end{equation}
\else
\begin{eqnarray}
H =  &-& t  \sum_{\langle ij\rangle} (c_i^\dagger c_j + c_j^\dagger c_i) \nonumber \\
       &+& \sum_i \varphi_i  n_i + \frac{1}{2}\sum_{i\not=j}(n_i-K)(n_j-K)U_{ij},
\label{eq:Hamiltonian}
\end{eqnarray}
\fi
where $c_i^\dagger$ and $c_i$ are the creation and annihilation operators
at site $i$, $n_i =c_i^\dagger  c_i$,  and $\langle ij \rangle$ denotes all pairs of nearest 
neighbors. $U_{ij} = e^2/r_{ij}$ represents the 
Coulomb interaction which is parametrized  by its nearest-neighbor value $U$
and $t$ is the kinetic energy.
The random potential values $\varphi_i$ are chosen 
from a box distribution of width $2 W$ and zero mean.
(We always set $W=1$.) Two important limiting cases of the quantum Coulomb glass are the Anderson model of
localization (for $U=0$) and the classical Coulomb glass (for $t=0$).

The numerical simulation of disordered many-particle systems is one of the most complicated 
problems in computational physics. First, the size of the 
Hilbert space grows exponentially with the system size making exact diagonalizations of
the Hamiltonian impossible already for very small systems. Second,  the presence
of disorder requires the simulation of many different samples to
obtain averages or distributions of physical quantities. For disordered interacting
electrons the problem is made worse by the long range of the Coulomb interaction
which has to be retained for a correct description of  the insulating phase
where screening breaks down.

In this Letter we suggest an efficient numerical method to simulate disordered interacting electrons.
It is based on the idea of configuration interaction \cite{fulde} adapted to disordered lattice models.
The method, which we call the Hartree-Fock based diagonalization (HFD),
consists of 3 steps: (i) solve the Hartree-Fock (HF) approximation of the Hamiltonian as in
Ref. \cite{epper_hf}, (ii) use a Monte-Carlo algorithm to find the low-energy many-particle
HF states (Slater determinants), (iii) diagonalize the Hamiltonian in
the basis formed by these states \cite{classical}. 
The HF basis states are comparatively close in character to the exact eigenstates in the entire
parameter space. Thus it is sufficient to keep only a small fraction of the Hilbert 
space to obtain low-energy  quantities with an accuracy comparable to that of exact 
diagonalization. The HFD method is very flexible, it works well in any spatial dimension, and is 
capable of handling long-range and short-range interactions. 
A detailed description will be given elsewhere.
Most of our calculations have been performed for  lattices with $5^2$ sites
and 12 electrons keeping 500 basis states.
We used periodic boundary conditions and the minimum image convention.
We also studied $4^2$ and $6^2$ systems with
 $K=0.25$ and 0.5 keeping up to 2000 out of $9*10^{9}$ basis states. 

We now turn to the conductance which we compute from linear response theory. 
The real (dissipative) part of the 
conductance (in units of $e^2/h$)
is given by the Kubo-Greenwood formula \cite{kubo_greenwood},
\begin{equation}
 {\rm Re} ~ G^{xx}(\omega) = \frac {2 \pi^2}  {\omega} \sum_{\nu} |\langle 0 | j^x|\nu \rangle |^2 
     \delta(\omega+E_0-E_{\nu})
\label{eq:kubo}
\end{equation}
where $\omega$ denotes the frequency. 
$j^x$ is the $x$ component of the current operator and $\nu$ denotes the eigenstates
of the Hamiltonian. Eq.\ (\ref{eq:kubo}) describes an isolated system while
 in a real d.c.\ transport experiment
the sample is connected to contacts and leads. This results in a finite life time $\tau$
of the eigenstates leading to an inhomogeneous broadening $\gamma = \tau^{-1}$
of the $\delta$ functions in (\ref{eq:kubo}) \cite{datta}. To suppress
the discreteness of the spectrum of a finite system, $\gamma$ should be 
at least of the order of the single-particle level spacing. For our systems this requires
a comparatively large $\gamma \ge 0.05$. We tested different $\gamma$
and found that the conductance {\em values} depend on $\gamma$ but the 
qualitative results do not \cite{gamma}. 

In a random system different samples 
will have different conductance values. Fig.\ \ref{fig:psig}
shows the probability distribution $P[\log(G(0))]$ 
for systems  in the localized regime with and without interactions. 
Both distributions show the same qualitative behavior, they are close to normal distributions 
corresponding to very broad distributions of the conductances themselves.
The arithmetic average of the conductance is therefore not a good measure of the 
typical behavior. We instead use the logarithmic (i.e. geometrical) average
$G_{typ} = \exp\langle\log (G )\rangle$ \cite{selfav}, usually over 400 disorder configurations. 
\begin{figure}
  \epsfxsize=\figuresize
  \centerline{\epsffile{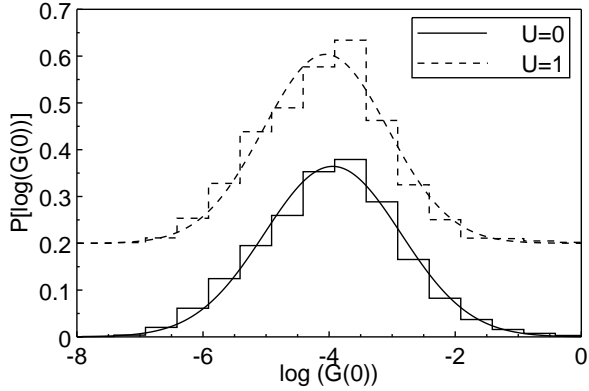}}
  \caption{$P[\log(G(0)]$ for  $W=1$, $t=0.1$, and $\gamma=0.05$. The histograms represent 2000 samples. 
              The smooth lines are fits to Gaussians. The data for $U=1$ have been shifted by 0.2}
  \label{fig:psig}
\end{figure}

In Figs.\ \ref{fig:sig_a} and \ref{fig:sig_b} we present results on the dependence 
of the conductance on the interaction for two sets of parameters. 
\begin{figure}
  \epsfxsize=\figuresize
  \centerline{\epsffile{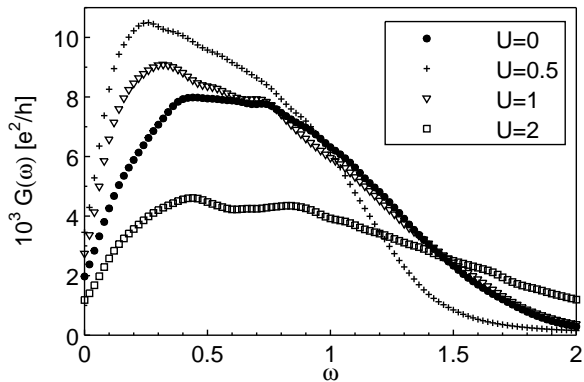}}
  \caption{$G(\omega)$ for   $W=1$,    $t=0.03$, $\gamma=0.05$.}
  \label{fig:sig_a}
\end{figure}
\begin{figure}
  \epsfxsize=\figuresize
  \centerline{\epsffile{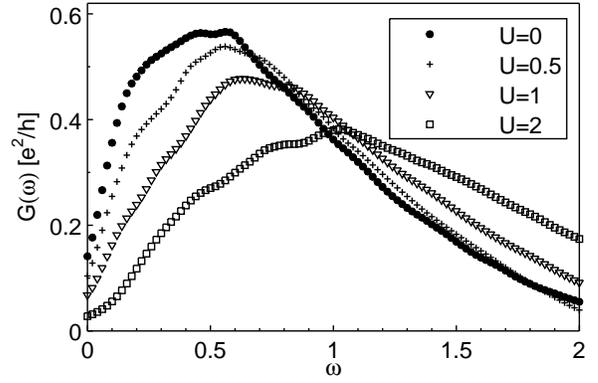}}
  \caption{Same as Fig.\ \protect\ref{fig:sig_a} but for $t=0.3$.}
  \label{fig:sig_b}
\end{figure}
In Fig.\ \ref{fig:sig_a} the kinetic
energy is very small ($t=0.03$). Thus the system is in the strongly localized regime,
as we also estimated from the single-particle participation number 
$P_{\rm sp} \approx 2$. Here a weak Coulomb interaction ($U=0.5$) leads to
an {\em increase} of the conductance at low frequencies. If the interaction 
becomes stronger the conductance decreases and finally ($U=2$) falls below the value of
non-interacting electrons. We emphasize that the increase of the conductance for weak
interactions is a true correlation effect: Within the HF approximation \cite{epper_hf} interactions
always lead to a decrease of the conductance. 
The behavior is qualitatively different at higher kinetic energy 
($t=0.3$) as shown in Fig.\ \ref{fig:sig_b}. Here
the system is approaching the diffusive regime ($P_{\rm sp} > 10$).
Already a weak interaction ($U=0.5$) leads to a reduction of the 
low-frequency conductance
compared to non-interacting electrons. If the interaction becomes stronger 
the conductance is decreased further.
We have performed analogous calculations for kinetic energies $t=0.01 ... 0.5$ and
interaction strengths $U=0 ... 2$. The resulting d.c.\ conductances are those presented
in Fig.\ \ref{fig:sigzero}. 

We also checked for system size and filling factor dependences
by simulating systems with $4^2$ and
$6^2$ sites and filling factor $K=0.25$ in addition to 0.5. We found the qualitative
picture  (as presented in Fig.\ \ref{fig:sigzero}) to be the same in all cases. As an example,
Fig.\ \ref{fig:g_l_t001} shows the interaction dependence of $G(0)$ for $t=0.01$ for the 
different systems studied. Clearly, the interaction induced enhancement of the conductance
exists in all cases. Moreover the relative enhancement seems to increase from the $4^2$ system to the 
$6^2$ system. (A comparison of
even and odd linear system sizes is problematic since at half filling  a regular array of charges is
impossible for odd sizes. 
Moreover, any {\em quantitative} comparison of different sizes would require 
a more realistic description of the broadening.)
\begin{figure}
  \epsfxsize=\figuresize
  \centerline{\epsffile{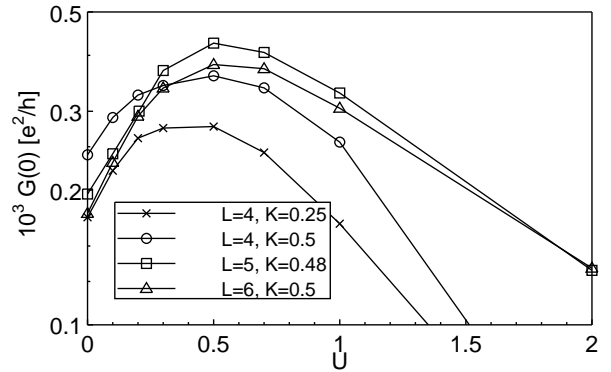}}
  \caption{Comparison of $G(0)$ for $W=1, t=0.01$ and different system sizes
        and filling factors.}
  \label{fig:g_l_t001}
\end{figure}

In order to find out to what extent the behavior of the conductance is reflected in single-particle
localization properties we also computed the single-particle return probability 
\begin{equation}
R_p(\varepsilon) = \frac 1 N \sum_j \lim_{\delta \to 0} \frac \delta \pi \, G_{jj}(\varepsilon + i \delta)
     \, G_{jj}(\varepsilon - i \delta) ~.
\end{equation}
Here $G_{ij}(\varepsilon)$ is the single-particle Greens function.
$R_p(\varepsilon)$ is the generalization of the inverse participation number 
$P_{\rm sp}^{-1}(\varepsilon)$ (of a single-electron state) to a many-particle system.
Fig.\ \ref{fig:return} shows a typical result for $R_p(\varepsilon)$.
\begin{figure}
  \epsfxsize=\figuresize
  \centerline{\epsffile{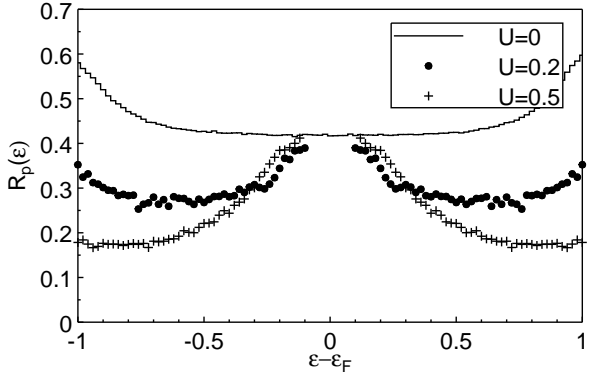}}
  \caption{$R_p(\varepsilon)$ for $W=1, t=0.1$. 
              The data are averaged over 2000 disorder configurations
              (10000 for non-interacting electrons).}
  \label{fig:return}
\end{figure}
We performed analogous calculations for $t=0.01 ... 0.5$ and $U=0 ... 2$.
For all cases  we obtain the same qualitative behavior: Close to the Fermi energy 
the return probability is only weakly influenced by the interaction. Directly at the 
Fermi energy, which is not accessible in our simulations because of our still too small 
system sizes, there may develop a slight enhancement of the return probability as a 
result of the Coulomb gap in the single-particle density of states. Such an enhancement 
has already been observed within the 
HF approximation  \cite{epper_hf}. Within the results obtained in this Letter, 
the effect, if any, is weaker than within HF. 
For energies away from the Fermi energy the single-particle excitations in the 
interacting system become strongly {\em de}localized compared to the non-interacting case.
The interaction dependence of the conductance discussed above is, 
however, not reflected in the single-particle return probability.

To summarize, we have used the Hartree-Fock based diagonalization (HFD) method
to investigate the transport properties of disordered interacting spinless
electrons. We have found that a weak Coulomb interaction can enhance the conductivity 
of localized samples considerably while it reduces the conductance in the case of weaker disorder.
If the interaction becomes stronger it eventually reduces the conductance also in the 
localized regime.
Let us finally mention that although we show that  interactions
can enhance the conductivity in certain parameter regions this does not directly
provide an explanation
for the MIT in 2D \cite{2DMIT} since the importance of the spin
degrees of freedom for this transition is established experimentally \cite{magnet}.
We emphasize however, that our method is very easy to generalize
to electrons with spin. Work in this direction is in progress.

We acknowledge financial support by the Deutsche Forschungsgemeinschaft.


\begin{references}
\bibitem{SO}except for the spin-orbit coupling universality class which allows for a MIT in 2D
       and quantizing magnetic fields in 2D where extended states can appear in the band center.
\bibitem{classical}A related method based on the low-energy states 
      of the classical Coulomb glass  was used before \cite{efros95,talamantes96}. 
     While a method based on the classical states is restricted to very small values of the 
     kinetic energy, our method works in the entire parameter space.
\bibitem{gamma}In order to compare the conductance values for different 
     kinetic energies, $\gamma$ should be $t$-dependent. In the diffusive regime a simple estimate
    \cite{datta} suggests $\gamma \propto t$. Since this would lead to unacceptable large
    broadenings for our systems we used a $t$-independent $\gamma$ in our calculations.
\bibitem{selfav}For noninteracting electrons it is known that $\log(G)$ is a statistically well-behaved 
   quantity, see e.g. \cite{kramrev93}.
\end{references}
\end{document}